\documentclass[twocolumn,showpacs,preprintnumbers,amsmath,amssymb]{revtex4}

\usepackage[ansinew,latin1]{inputenc}
\usepackage{amsmath}
\usepackage{amssymb}
\usepackage{graphicx} 

\def\Journal#1#2#3#4{{#1} {\bf #2}, #3 (#4)}

\def\PLB{{Phys. Lett.}  B}
\def\PRL{Phys. Rev. Lett.}
\def\PR{Phys. Rev.}
\def\PRC{{Phys. Rev.} C}

\begin{document}

\title{Neutron scattering and extra-short-range interactions}

\author{V.~V.~Nesvizhevsky}
	\email{nesvizhevsky@ill.eu}
  \affiliation{Institute Laue Langevin, 6 rue Jules Horowitz, F-38042, Grenoble, France}
\author{G.~Pignol}
  	\email{pignol@lpsc.in2p3.fr}
\author{K.~V.~Protasov}
	\email{protasov@lpsc.in2p3.fr}
 \affiliation{Laboratoire de Physique Subatomique et de Cosmologie,
 UJF - CNRS/IN2P3 - INPG, 53 Av. des Martyrs, Grenoble, France}

\date{\today}

\begin{abstract}
The available data on neutron scattering were analysed to constrain a
hypothetical new short-range interaction.
We show that these constraints are several orders of magnitude better than
those usually cited in the range between 1~pm and 5~nm.
This distance range occupies an intermediate space between collider
searches for strongly coupled heavy bosons and searches for new weak
macroscopic forces.
We emphasise the reliability of the neutron contraints in so far as they
provide several independent strategies.
We have identified the most promising way to improve them.
\end{abstract}

\maketitle

\section{Introduction}

The existence of other forces in nature, mediated by new bosons, has been
extensively discussed in the literature, given their possibility in many of
the extensions to the standard model of particle physics \cite{PDG}.
New bosons for example are predicted by most of the Grand Unified Theories
embedding the standard model, with a coupling constant of $\approx 10^{-1}$.
These strongly coupled bosons would have to be heavier than $\approx 1$
TeV if they were not to conflict with present observations; 
heavier bosons will be searched for at the Large Hadron Collider.
Lighter bosons could however have remained unnoticed, provided they
interact weakly with matter.
Such bosons would mediate a finite range force between two fermions:
\begin{equation}
\label{Yukawa}
V(r) = Q_1 Q_2 \ \frac{g^2}{4 \pi} \frac{\hbar c}{r} e^{-r/\lambda}
\end{equation}
where $g$ is the coupling constant, $Q_1$ and $Q_2$ the charges of the
fermions under the new interaction, and the range of this Yukawa-like
potential $\lambda = \frac{\hbar}{M c}$ is inversely proportional to the
boson mass $M$. 
In the following we consider the interactions of neutrons with nuclei of
atomic number $A$: the charge of the atom under the new interaction is
equal $Q_1=A$; the neutron charge is equal unity $Q_2=1$. 
A new boson could even be massless, as has been suggested by Lee and Yang
\cite{Lee:1955vk}
well before the birth of the standard model, to explain the conservation
of the baryon number.
This additional massless boson would mediate a new infinite-range force,
and could be seen in searches for violation of the equivalence principle.
The presence of very light bosons ($M \ll 1$ eV) would be shown by
deviations from the gravitational inverse square law.
Gravity has been probed down to distances of $0.1$~mm
\cite{Kapner:2006si}; new bosons lighter than $2 \times 10^{-3}$~eV 
must thus have a coupling constant lower than the gravity strength between
nucleons, $g^2 < 10^{-37}$.

Theories with extra large spatial dimensions
\cite{Arkani-Hamed:1998rs,Antoniadis:1998ig,Rubakov:1983bb,Visser:1985qm,Antoniadis:1990ew,Lykken:1996fj} 
provide strong motivation to search for such forces. 
If a boson is allowed to travel in large extra-dimensions, with a strong
coupling constant in the bulk, it behaves in our 4D world as a very weakly
coupled new boson, the coupling being diluted in the extra-dimensions.
The light dark matter hypothesis also argues in favour of the existence of
new short range interactions \cite{Fayet:2007ua}.

While measurements of the Casimir or Van der Waals forces (for a review,
see e.g. \cite{Bordag:2001qi}) give the best constraints in the nanometer
range ($10 \ \rm{nm} < \lambda < 1 \ \mu \rm{m}$), 
and antiprotonic atoms constrain the domain below 1 pm \cite{ASAKUZA,
Nesvizhevsky:2004qb}, 
it has been suggested that experiments with neutrons could be competitive
in the intermediate range \cite{Leeb:1992qf, Frank:2003ms, Watson:2004vh,
Greene:2006qj, Nesvizhevsky:2004qb, Nesvizhevsky:2007fv}.
Neutrons could also probe spin-dependent interactions in a wider distance
range \cite{Baessler:2006vm}, 
or spin-independent interactions in the range of several micrometers
\cite{Nesvizhevsky:2007fv,Nesvizhevsky:2002ef}.

In this article we give the quantitative constraints on the parameters of
the additional interaction, $\lambda$ and $g$, 
from the existing data on neutrons scattering at nuclei.
In section \ref{sect1} we analyse the influence of a new short-range
interaction on the scattering of neutrons at nuclei.
In section \ref{Adep}, we use the fact that the nuclear radius, as well as
the scattering lengths, are expected to be proportional to $A^{1/3}$, where
$A$ is the number of nucleons, 
whilst the contribution of an additional interaction would result in an
additional linear term in the mass dependence of the scattering lengths.
In section \ref{sect2} and \ref{sect3}, we use the different sensitivities of different
types of neutron scattering experiments to extra interactions (forward and
backward scattering) in order to constrain them.
In section \ref{proposal} we propose a way to improve these constraints
\cite{Nesvizhevsky:2007fv}.

\section{Slow neutron / nuclei interaction with extra-short-range interactions}
\label{sect1}

The scattering of slow neutrons on atoms is described by the scattering amplitude 
$f(\mathbf{q})$; this can be represented by a sum of a few terms \cite{Sears}:
\begin{equation}
\label{amplitude_tot}
f(\mathbf{q}) = f_{\mbox{\tiny nucl}}(\mathbf{q}) + f_{ne}(\mathbf{q})+ f_V(\mathbf{q})
\end{equation}
The first and the most important term represents the scattering due to 
the nuclear neutron-nucleus interaction. 
At low energies discussed in this article, it is isotropic and energy-independant, 
because the nuclear radius is much smaller than the wavelegth of slow neutrons:
\begin{equation}
\label{bnuc}
f_{\mbox{\tiny nucl}}(\mathbf{q}) = - b.
\end{equation}
The coherent scattering lenght $b$ is the fundamental parameter 
describing the interaction of slow neutrons with a nucleus \cite{Fermi}.

The second term is the amplitude of so-called electron-neutron scattering 
due to the interaction of the neutron charge distribution with
the nucleus charge and the electron cloud. 
This amplitude can be written as
\begin{equation}
\label{bne}
f_{ne}(\mathbf{q}) = - b_{ne}(Z-f(Z,\mathbf{q})),
\end{equation}
where $f(Z,\mathbf{q})$ is the atomic form-factor measured in the X-rays experiments 
and $b_{ne}$ is a constant called the electron-neutron scattering length, 
which is directly related to the neutron charge radius \cite{Sears} and to the neutron electromagnetic
form-factor $G_E(\mathbf{q}^2)$ by
\begin{equation}
\label{radius}
b_{ne}= - \left. \frac{2}{a_0} \frac{m}{m_e} \frac{dG_E(\mathbf{q}^2)}{d \mathbf{q}^2}\right|_{\mathbf{q}^2 = 0},
\end{equation}
$m$ and $m_e$ being the neutron and electron masses, $a_0$ the Bohr radius.
This contribution to the total scattering amplitude is as small as a per cent for heavy nuclei.

In the presence of a new interaction (\ref{Yukawa}), the scattering 
for a center of mass momentum $\hbar k$ due to the extra interaction,
within the Born approximation, is given by
\begin{equation}
\label{amplitude}
f_V(\theta) = - A \frac{g^2}{4 \pi} \hbar c \frac{2 m \lambda^2/\hbar^2}{1+ \left(q \lambda \right)^2}
\end{equation}
where $q = 2 k  \sin(\theta/2)$, $\theta$ is the scattering angle.

Any other possible contributions to the scattering amplitude 
$f(\mathbf{q})$, due to non zero nuclear radius, nucleon polarizability, etc.
are very small in the energy range discussed here \cite{Sears} 
and have therefore been omitted in (\ref{amplitude_tot}).

The nuclear scattering lengths are measured for almost all stable nuclei,
using a variety of methods.
A review of the different methods and a complete table of the measured
scattering lengths can be found in \cite{Rauch}.
We can distinguish two classes of method, with different sensitivities to
a new interaction.

The first class -- including the interference method, the total reflection method, 
the gravity refractometer method -- measures the forward scattering amplitude
$f(\mathbf{q}=0)$.
These methods actually measure the mean optical potential of a given material, 
called the Fermi potential, due to the coherent scattering of neutrons at many nuclei.
The Fermi potential is related to the {\sl forward} scattering amplitude.

In the presence of the new force, the measured scattering lenght 
can be separated into a nuclear and an additional term\footnote{For simplicity,
the amplitude due to electron-neutron scattering has been omitted for the
time being. This term will be given detailed treatment in section
\ref{sect3}.}:
\begin{equation}
\label{Opt}
b_{\mbox{\tiny opt}} = - f(\mathbf{q}=0) = b + A \frac{m c^2}{2 \pi \hbar c} \ g^2 \lambda^2
\end{equation}
The second class of method -- including the Bragg diffraction method and
the transmission method -- uses non-zero transferred momentum.
In the Bragg diffraction method, the scattering amplitude for a momentum 
transfer of $q_{\mbox{\tiny BD}} = 10 \ {\rm nm}^{-1}$ is measured.
One actually extracts, besides the nuclear term, an extra contribution according to (\ref{amplitude}) 
\begin{equation}
\label{BD}
b_{\mbox{\tiny BD}} = b + A \frac{m c^2}{2 \pi \hbar c} \ g^2  \frac{\lambda^2}{1 + \left( q_{\mbox{\tiny BD}} \lambda \right)^2}
\end{equation}
In the case of the transmission method, the total cross-section is measured. 
Generally, neutrons with energies of about $1$ eV are used; they are much
faster than slow neutrons, and no coherent scattering can be observed.
An additional interaction would manifest itself by an energy dependance of
the extracted scattering length
\begin{equation}
\label{TR}
b_{\mbox{\tiny TR}}(k^2) = \sqrt{\frac{\sigma_{\rm tot}}{4 \pi}} = b + A \frac{m c^2}{2 \pi \hbar c} \ g^2 \lambda^2  \frac{\ln ( 1+ 4 (k \lambda)^2) }{4 (k \lambda)^2}
\end{equation}
Finally, we should also mention the very popular Christiansen filter
technique; this measures \emph{relative} scattering lengths, so we do not
consider this data.

\section{Random potential nuclear model}
\label{Adep}

\begin{figure}
\begin{center}
\includegraphics[width=.95\linewidth]{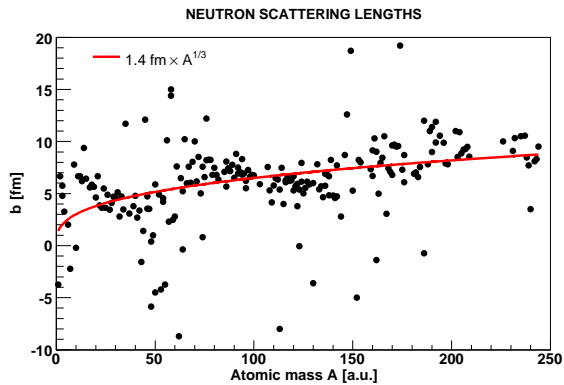}
\end{center}
\caption{
Measured scattering lengths as a function of nucleus atomic number.
} 
\label{scattering_lengths}
\end{figure}

A simple and robust limit on the additional Yukawa forces can be easily 
obtained by neglecting the small term due to
the neutron-electron scattering and by studing the general $A$-dependence 
of the scattering amplitude. 
In the domain of $\lambda \leq 1/q_{\mbox{\tiny BD}}$, 
the optical and Bragg diffraction methods are sensitive to the same
amplitude
\begin{equation}
b_{\rm Meas} = - f(\mathbf{q}=0) = b + A \frac{m c^2}{2 \pi \hbar c} \ g^2 \lambda^2
\end{equation}
as clear from (\ref{Opt}) and (\ref{BD}). 
The presence of additional forces would be apparent from the linear
increase of the measured scattering length as a function of $A$. 
This dependence is not observed in the experimental data
presented in fig.~\ref{scattering_lengths}. 
Strong variations with $A$ of the measured scattering length are mostly due to the
shell effects in neutron-nucleus nuclear amplitude $b$.

To establish a quantitative upper limit on this additional term, the $A$
dependence of the nuclear scattering length $b=b(A)$ needs to be taken
into account. 
The complete {\sl ab initio} calculation of these shell
effects for $b$ is particularly complicated and has never been reported in
the literature.  
Fortunately, there exists a very simple and elegant semi-phenomenological
approach that describes these variations \cite{Peshkin}.
It assumes that a nucleus can be presented as an attractive "square well"
potential, 
with radius $R A^{1/3}$ and depth $V_0$ for slow neutrons.
The scattering length would then be equal to
\begin{equation}
b(A) = R A^{1/3} \left( 1 - \frac{\tan(X)}{X} \right),
\end{equation}
where $X = \frac{R A^{1/3}}{\hbar} \sqrt{2 m V_0}$ is supposed to be a
random variable distributed uniformly over the range $\left[ \pi/2, 5 \pi/2
\right]$~; 
the lower value corresponds to the appearance of a bound state and the
upper limit is set sufficiently large not to influence the results of the
present analysis; more details can be found in \cite{Peshkin}.

\begin{figure}
\begin{center}
\includegraphics[width=.95\linewidth]{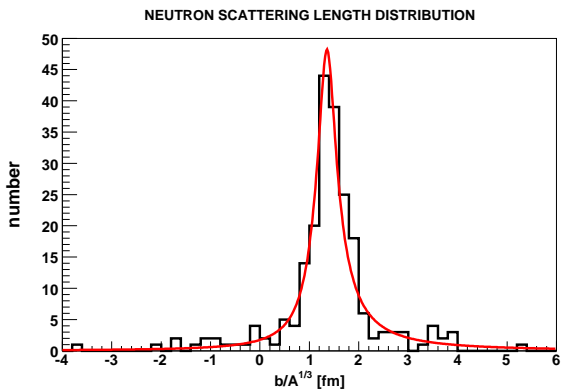}
\end{center}
\caption{
This histogram shows the distribution of measured scattering lengths
normalized to the radius of the nuclei.
The curve corresponds to the random potential model.
} 
\label{modeldependant}
\end{figure}

\begin{figure}
\begin{center}
\includegraphics[width=.95\linewidth]{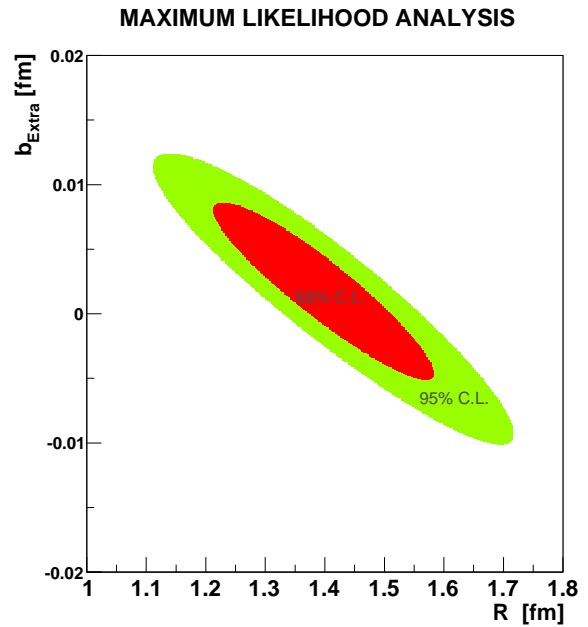}
\end{center}
\caption{
Maximum likelihood analysis of the two parameters $R$ and $b_{\rm Extra}$ 
of (\ref{nuclearpluslinearterm}).
} 
\label{likelihood}
\end{figure}

Since 1970 (\cite{Peshkin}), the number of nuclei with measured scattering
lengths has doubled and the level of precision has been improved.
We include here the new interaction analysis.

Without any additional interaction, this analysis provides the results presented in fig.~\ref{modeldependant}.
One may conclude that the model describes well the distribution of all experimental data well;
the value of the only free parameter in this model is estimated to be $R = 1.44 \pm 0.05$~fm at the 68 \% C.L.
The likelihood function at its maximum satisfies $\ln(L) = -254$ for 216 degrees of freedom.

With a short-range new interaction included in the analysis 
we have to consider the random variable
\begin{equation}
\label{nuclearpluslinearterm}
b_{\rm Meas} =  R A^{1/3} \left( 1 - \frac{\tan(X)}{X} \right) + b_{\rm Extra} \ A.
\end{equation}
where the effect of the extra interaction is the slope 
$b_{\rm Extra} = \frac{m c^2}{2 \pi \hbar c} \ g^2 \lambda^2$ of the linear term.
The estimation of the two parameters $R$ and $b_{\rm Extra}$ 
from the experimental data, again using the maximum likelihood method, is presented in fig.~\ref{likelihood}.
The linear term is compatible with zero, as expected.
We thus obtain a quantitative constraint for the coupling $g(\lambda)$:
\begin{equation}
g^2 \lambda^2 \leq 0.016 \ {\rm fm}^2 \quad {\rm at \ 95 \% \ C.L.}
\end{equation}
This result is presented in fig.~\ref{exclusion} for the distance range of interest, $10^{-12} - 10^{-10}$~m.

\section{Constraint from comparaison of forward and backward scattering of neutrons}
\label{sect2}

Another way to constrain on aditional Yukawa forces consists 
in comparing the scattering lengths measured by different methods.

As explained above, the scattering lengths measured using the Bragg
diffraction method $b_{\mbox{\tiny BD}}$ 
and the interference method $b_{\mbox{\tiny opt}}$ do not show the same
sensitivity to a new short-range interaction.
According to (\ref{Opt}) and (\ref{BD}), the ratio of the two values
should deviate from unity in the presence of an additional interaction
\begin{equation}
\label{BraggVSInterf}
\frac{b_{\mbox{\tiny opt}}}{b_{\mbox{\tiny BD}}} \approx 1 + \frac{A}{b} \
\frac{m c^2}{2 \pi \hbar c} \ g^2 \ \lambda^2 \ \frac{(q \lambda)^2}{1+(q
\lambda)^2}
\end{equation}
We found a set of 13 nuclei for which both measurements exist.

The different measurements quoted in the litterature are sometimes incompatible, 
even for the same measurement method, 
because of additional systematics not included in the quoted errors.
To compensate for this we estimated a 
\emph{methodological error} $\sigma$ for each method.
By selecting all the nuclei for which several measurements with a given method are available, 
we obtained the values $b_{A, i} \pm \Delta b_{A, i}$, 
where $A$ is the nucleus index and $1 \leqslant i \leqslant n_A$ 
lists the experiments using the given method.
The methodological error is then calculated so that:
\begin{equation}
\sum_{A, i} \frac{\left( b_{A, i} - \bar{b}_A \right)^2}{\sigma_{A, i}^2 + \sigma^2} = \sum_A n_A.
\end{equation} 
That is, we force the $\chi^2$ deviation from the weighted average $\bar{b}_A$ 
to be equal to the number of degrees of freedom $\sum_A n_A$.
For the Bragg diffraction method, we have 72 degrees of freedom, 
and a methodological error of $0.16$ fm; 
this can be compared to the average of the publised errors of $0.23$ fm.
For the interference method, we have 22 degrees of freedom; 
we found a methodological error of $0.05$~fm, 
while the average of published errors is $0.05$~fm.
Table \ref{methods} summarizes.

With the enlarged errors, a $\chi^2$ fit of eq.~(\ref{BraggVSInterf}) 
using the 13 set of data for $b_{\rm I}$ and $b_{\rm BD}$, we obtain the constraint: 
\begin{equation}
g^2 \lambda^2 \frac{(q \lambda)^2}{1+(q \lambda)^2} \leq 0.0013 \ {\rm fm}^2 \quad {\rm at \ 95 \% \ C.L.}
\end{equation}
corresponding to the bold limit in fig.~\ref{exclusion}.

\begin{table}
\begin{center}
\caption{\label{methods} Existing data on neutron scattering lengths.}
\begin{tabular}{lcc}
Method	&  {\small Bragg} & {\small Interference} \\
  	& {\small Diffraction}  \\
\hline
Number of measurements		& 141	& 41	 \\
Number of measured nuclei	& 98	& 28	 \\
Mean relative accuracy		& 3 \%	& 0.4 \% \\
Methodological error [fm]	& 0.16	& 0.05 
\end{tabular}
\end{center}
\end{table}

\section{Electromagnetic effects}
\label{sect3}

Up to now, the amplitude due to a new additional interaction
$f_V(\mathbf{q})$ has been compared to the nuclear one
$f_{\mbox{\tiny nucl}}(\mathbf{q})$ (see (\ref{amplitude_tot})).
One could improve the limit obtained by comparing the amplitude due to a
new additional
interaction to a smaller amplitude due to an electromagnetic interaction
($f_{ne}(\mathbf{q})$).
This idea was first proposed in ref. \cite{Leeb:1992qf}.

To achieve this, three independent measurements are required 
(roughly speaking, to determine independently the
three contributions to (\ref{amplitude_tot})).

One could repeat the previous analysis using measurements
of the total cross-section instead of the Bragg diffraction. 
As can be seen from (\ref{TR}), the scattering length extracted at
energies of $\approx 1$~eV ($k = 200 \ {\rm nm}^{-1} = 1 / 5 \ {\rm pm}$) contains an extra force contribution that is
different from that in optical methods. 
In particular, if the range of a new interaction is larger than $1$~pm, 
the scattering length extracted from the total cross-section at $1$~eV can
be considered free of any extra contribution.
The sensitivity is so high, however, that the residual electromagnetic
effects due to the neutron square charge radius can mimick 
an extra-force contribution in the quantity $b(1 \ {\rm eV}) - b_{\rm
opt}$, as this contribution is energy-dependent and proportional to the
charge number of the atoms.
This problem is known as the determination of the neutron-electron
scattering length $b_{ne}$.
The extracted difference $b(1 \ {\rm eV}) - b_{\rm opt}$ therefore
contains the two contributions:
\begin{equation}
b(1 \ {\rm eV}) - b(0) = Z b_{ne} - A \frac{m c^2}{2 \pi \hbar c} \ g^2
\lambda^2 \left( 1 - \frac{\ln ( 1+ 4 (\frac{\lambda}{5 \ {\rm pm}})^2) }{4
(\frac{\lambda}{5 \ {\rm pm}})^2}\right)
\end{equation}

Unfortunately, there is very clear disagreement between the two groups of
values for $b_{ne}^{\rm exp} = \frac{b(1 \ {\rm eV}) - b(0)}{Z}$ known as the Garching-Argonne and Dubna values  \cite{Kopecky}
\begin{eqnarray}
\nonumber
b_{ne}^{\rm exp} & = & (-1.31 \pm 0.03) \times 10^{-3} \ {\rm fm} \ [\mbox{Gartching-Argonne}]\\
b_{ne}^{\rm exp} & = & (-1.59 \pm 0.04) \times 10^{-3} \ {\rm fm} \ [\mbox{Dubna}]
\end{eqnarray}
The discrepancy is much greater than the quoted uncertainties of the
experiments and there evidently an unaccounted for systematic error in at
least one of the experiments.

In order to overcome this difficulty we could determine $b_{ne}$ from the
experimental data on the
neutron form factor (\ref{radius}). 
The simplest way to do this consists in using a commonly accepted general
parametrization of the neutron form factor \cite{Glaster}:
\begin{eqnarray}
\label{GE}
G_E(\mathbf{q}^2)= -a \mu_n  \frac{\tau}{1+b \tau} G_D ,
\end{eqnarray}
where $\mu_n= - 1.91 \mu_B$ is the neutron anomalous magnetic moment, $\tau=\mathbf{q}^2/4m^2$ and 
\begin{eqnarray}
G_D(\mathbf{q}^2)= \frac{1}{\left(1+\mathbf{q}^2 / 0.71 \ ({\rm GeV/c})^2\right)^2} ,
\end{eqnarray}
is so-called dipole form factor ; $a$ and $b$ being fitting parameters.

\begin{figure}
\begin{center}
\includegraphics[width=.95\linewidth]{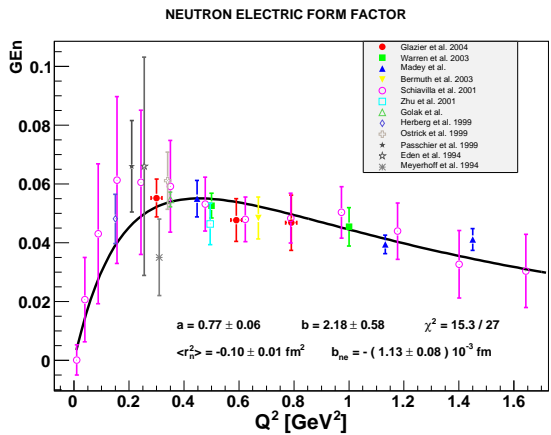}
\end{center}
\caption{
The neutron form factor $G_E(\mathbf{q}^2)$ as a function of the momentum transferred $\mathbf{q}^2$.
The experimental data are taken from \cite{FF}; 
the solid curve is a two parameter fit using formula (\ref{GE}).} 
\label{FF}
\end{figure}

A fit of an existing set of the neutron form factor experimental data \cite{FF} 
yields the following values for the parameters:
\begin{eqnarray*}
a & = & (0.77 \pm 0.06)  \\
b & = & (2.18 \pm 0.58)  
\end{eqnarray*}
with $\chi^2/{\mbox{NDF}}= 15.3/27$. 
The results of the fit are presented in  fig.~\ref{FF}.

Let us note that the momentum transferred $\mathbf{q}^2$ in these experiments 
is very large and a contribution from the term $f_V(\mathbf{q})$ is negligible.
The $b_{ne}$ determined in this way is
\begin{equation}
b_{ne} = \left( - 1.13 \pm 0.08 \right) \times 10^{-3} \ {\rm fm}.
\end{equation}
It does not agree with the value of $b_{ne}$, obtained in Dubna experiments.

Our principal conclusion consists in the observation of (underestimated)
systematical uncertainties in the presented experiments.
Therefore a single experiment/method can not be used for any reliable constraint.
A conservative estimate of the precision of the $b_{ne}$ value could be
obtained from analysing the discrepancies in the results obtained by
different methods; 
it is equal to $\Delta b_{ne} \leqslant 6 \times 10^{-4} \ {\rm fm}$.
The corresponding contraint at the $2 \sigma$ level
\begin{equation}
\frac{m c^2}{2 \pi \hbar c} \ g^2 \lambda^2 \left( 1 - \frac{\ln ( 1+ 4
(\frac{\lambda}{5 \ {\rm pm}})^2) }{4 ( \frac{\lambda}{5 \ {\rm
pm}})^2}\right) \leqslant \Delta b_{ne}
\end{equation}
is represented by the dot-dashed line in fig.~\ref{exclusion}.

\section{Asymmetry of scattering}
\label{proposal}

\begin{figure}
\begin{center}
\includegraphics[width=.95\linewidth]{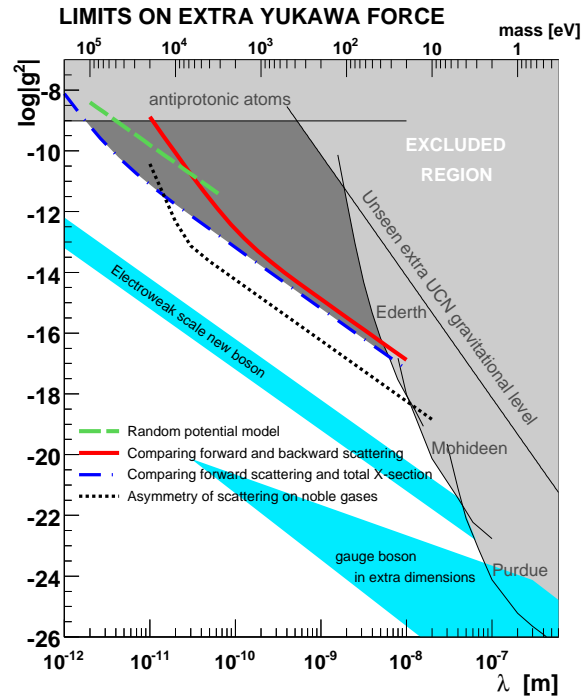}
\end{center}
\caption{
Constraint on extra Yukawa force at 95 \% C.L. (dashed, dot-dashed and bold lines) obtained in this article.
The dotted line is an estimation of the sensitivity of the proposed experiment.
Also shown the existing contraints \cite{Bordag:2001qi, Nesvizhevsky:2004qb} and the theoretical regions of interest \cite{Arkani-Hamed:1998rs,Fayet:2007ua}.
} 
\label{exclusion}
\end{figure}

As is clear from fig.~\ref{exclusion}, the best constraint was obtained 
from the analysis of the energy dependence of the neutron scattering
lengths in the $b_{ne}$ measurements. 
However, the precision here is limited by the correction for the $b_{ne}$
value itself.
An obvious proposal for improving this constraint would be to set up
experimental conditions free of the $b_{ne}$ contribution.
This is indeed possible, because neutron-electron scattering is essential
for fast neutrons only, and is absent for slow neutrons.

We propose improving the experiment \cite{Krohn} and measuring the
forward-backward asymmetry of the scattering of neutrons at atoms of noble
gases, in the following way: 
the initial velocity of the neutrons should correspond to the range of
very cold neutrons (VCN); 
the double differential measurement of neutron velocity before/after
scattering should be used to calculate the transferred momentum for every
collision.

\begin{figure}
\begin{center}
\includegraphics[width=.95\linewidth]{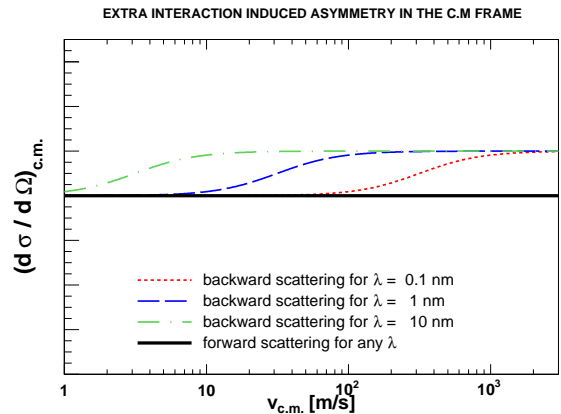}
\end{center}
\caption{
The asymmetry of neutron scattering at an atom in the center-of-mass reference system 
is shown as a function of various characteristic ranges 
of the additional interaction $0,1; 1; 10$ nm.} 
\label{asymmetry}
\end{figure}

\begin{figure}
\begin{center}
\includegraphics[width=.49\linewidth]{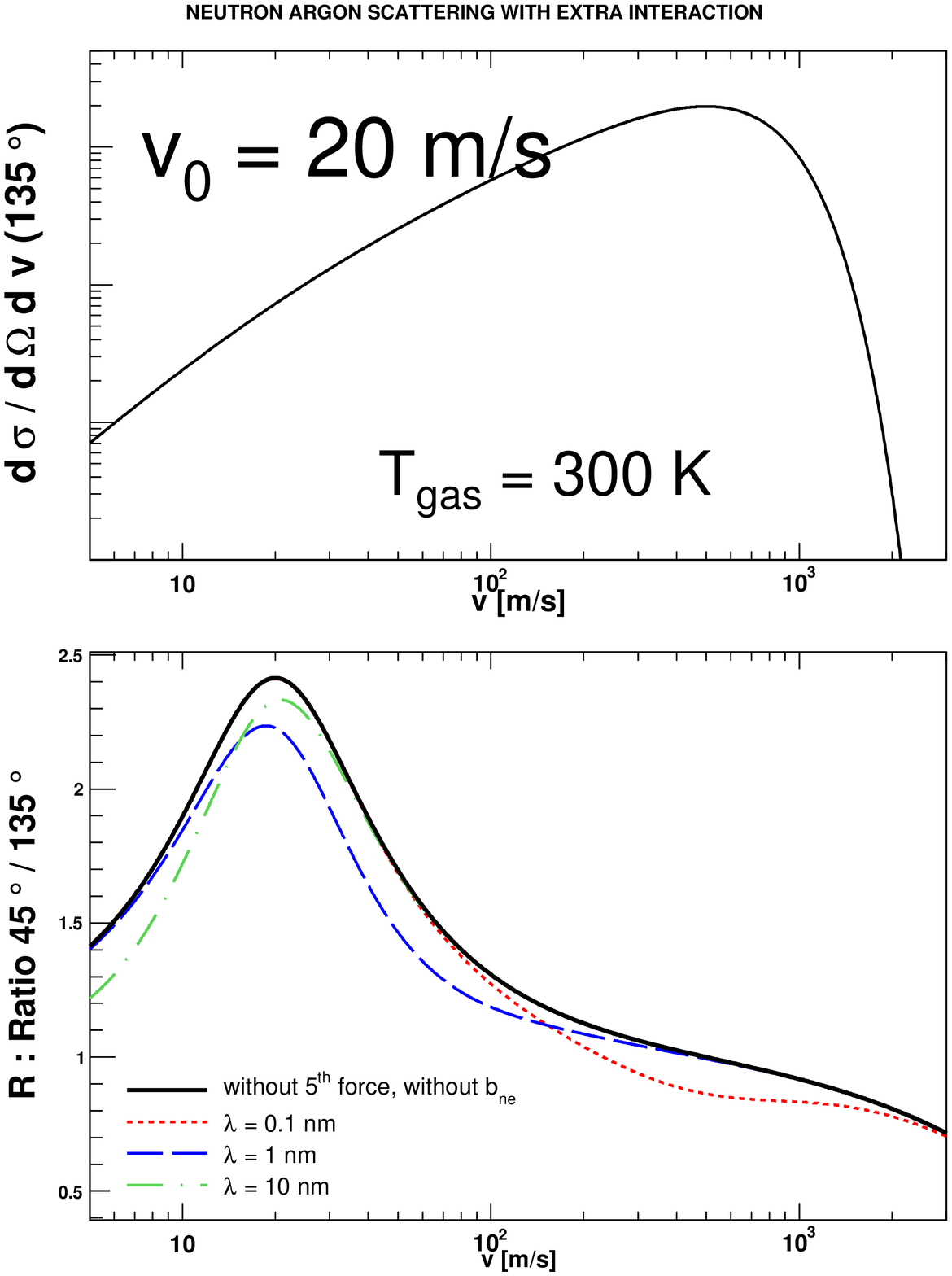}
\includegraphics[width=.49\linewidth]{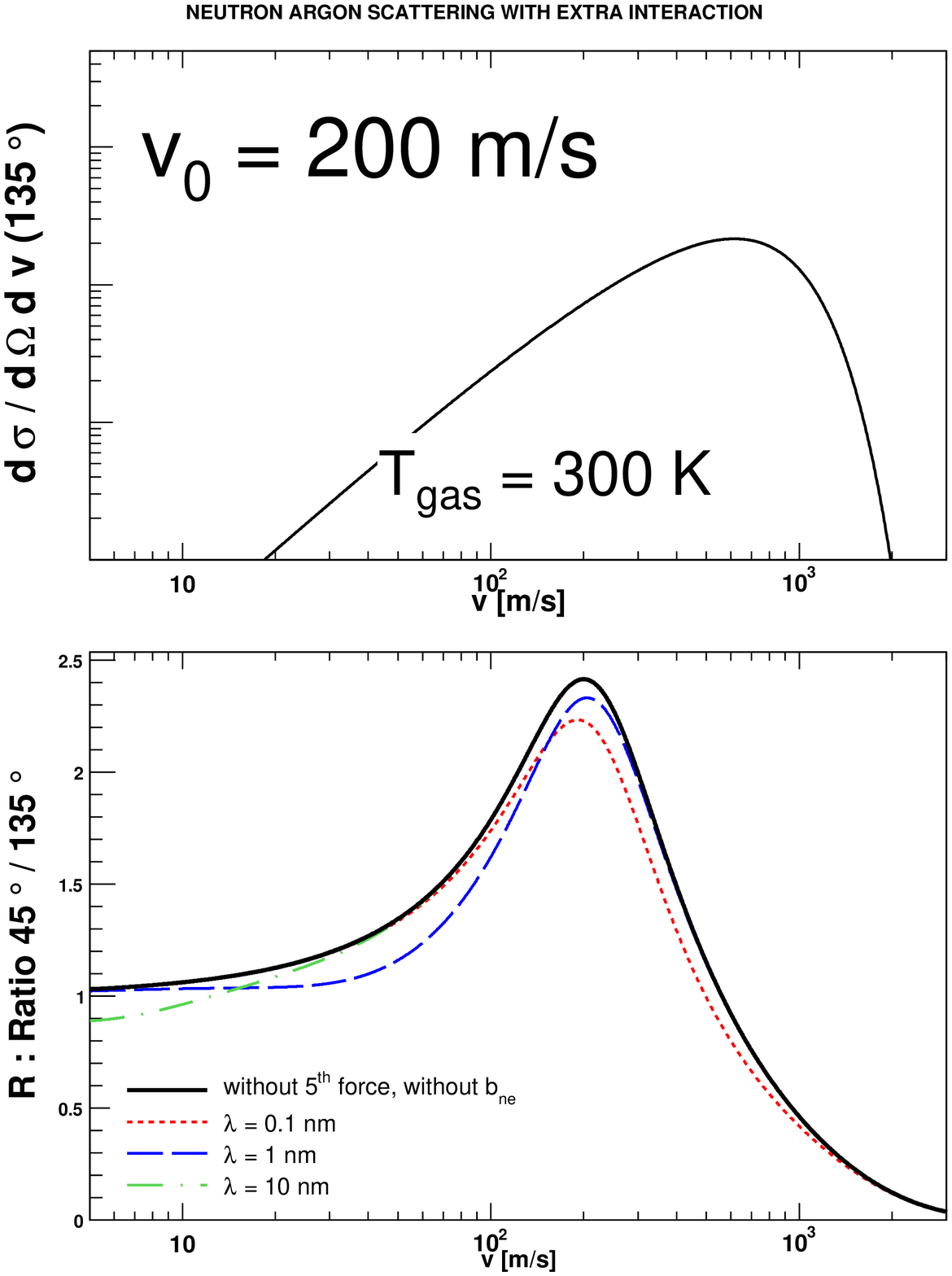}
\end{center}
\caption{
The bottom two graphs show the ratio of neutron flux scattered at an argon
atom at an angle of 45 degrees (forward) to that scattered at an angle of
135 degree (backward), as a function of the final neutron velocity for the
following cases: no additional interactions, 
additional interactions with characteristic ranges of $0.1; 1; 10$ nm. 
The thermal motion of gas is taken into account. 
The top two graphs indicate the rate of collision with argon atoms (in
thermal motion) of neutrons 
with a fixed initial neutron velocity.
The initial neutron velocity for the two graphs on the left is equal to
$20$ m/s; 
the initial neutron velocity for the two graphs on the right is equal to
$200$ m/s.
} 
\label{thermal}
\end{figure}

Fig.~\ref{asymmetry} shows the asymmetry of neutron scattering at an atom
in the center-of-mass reference system, 
as a function of various characteristic ranges of the additional
interaction. 
It is clear that neutrons with velocities of $1 - 1000$~m/s need to be
used.  
This means that the thermal motion of atoms cannot be neglected, and that
the neutron velocity in the laboratory fixed reference system is not equal
to that in the center-of-mass reference system. 
Even totally isotropic scattering in the center-of-mass reference system
would be highly anisotropic in the laboratory system. 
Nevertheless, the kinematics of the scattering process can be
reconstructed precisely if both the initial and the final neutron velocity 
in the laboratory reference system are measured. 
This can easily be done with reasonable statistical accuracy 
for thermal, cold, very cold and probably even for ultracold neutrons. 
In such a case, the asymmetry calculated would be modified by an
additional short-range interaction as shown in fig.~\ref{thermal} 
(the thermal motion of argon atoms is taken into account). 

The measurement described above could easily provide an accuracy of
$10^{-3}$ for the ratio of forward to backward scattering probabilities and
a corresponding constraint for the additional short-range interactions
shown in fig.~\ref{exclusion}.  
The relative drop in sensitivity at a few times $10^{-11}$ m is due to the
appearance of neutron electron scattering; 
the range of interest for this possible constraint is $10^{-11} - 10^{-8}
\ {\rm m}$.

\section{Conclusion}

We analysed the constraints for extra short-range interactions on the
basis of the existing data on neutron scattering.
These constraints are several orders of magnitude better than those
usually cited in the range between 1~pm and 5~nm.
The reliability of these constraints was supported by the application of
several independant methods with comparable accuracy, as well as by the use
of a major fraction of known neutron scattering lengths and treatment of
the data in a most conservative way.
One constraint obtained within the random potential nuclear model was
based on the absence of an additional linear term in the mass dependance of
the neutron scattering lengths.
It would be difficult to improve this constraint in either experimental or
theoretical terms.
Another constraint was derived by comparing two types of neutron
scattering experiments with different sensitivities to the extra
short-range interactions. 
These are interference experiments measuring forward neutron scattering
and the Bragg diffraction.
The accuracy here is limited by the relatively poor precision of the Bragg
scattering technique.
Significant improvements in the accuracy of such experiments would be
particularly interesting. 
Further constraints were estimated using the energy-dependence of the
neutron scattering lengths at heavy nuclei. 
They are limited by the precision of our knowledge of the neutron-electron
scattering length. 
An elegant method for further improving such constraints would consist in
achieving experimental conditions free of $b_{ne}$ contribution. 
This is indeed possible, given that neutron-electron scattering is
essential for fast neutrons only. 
The experiment would consist in scattering very cold neutrons at rare
noble gases and in measuring precisely the differential asymmetry of such
scattering as a function of the transferred momentum.

\section*{Acknowledgement}

We are grateful for a number of stimulating discussions with our
colleagues from the GRANIT project and the participants of the GRANIT 2006
workshop. 
This work is supported by the French Agence Nationale de la Recherche (ANR).


\end{document}